# Palladium-Coated Laterally Vibrating Resonators (LVRs) for Hydrogen Sensing

Gaia Giubilei, Farah Ben Ayed, Yvonne Sautriot, Aurelio Venditti, Kun Zhang,
Sila Deniz Calisgan, Pietro Simeoni, Zhenyun Qian, and Matteo Rinaldi
Institute for NanoSystems Innovation (NanoSI), Northeastern University, Boston, MA, USA

*Abstract*—This work presents a novel hydrogen sensor based on 30 % scandium-doped aluminum nitride (ScAlN) laterally vibrating resonators (LVRs) functionalized with a palladium (Pd) thin film. The micro-electro-mechanical system (MEMS) device operates by detecting shifts in resonant frequency resulting from hydrogen absorption in the Pd layer. The sensor demonstrates a high mechanical quality factor ($Q_m$) of 820, an electromechanical coupling coefficient ($k_t^2$) of 3.18 %, and an enhanced responsivity of 26 Hz/ppm in the low-parts-per-million (ppm) range, making it highly suitable for hydrogen leak detection. Compared to existing MHz-range technologies, the sensor achieves up to 50× higher sensitivity, while also offering multi-frequency definition in a single lithographic step, minimal footprint, and the highest quality factor among comparable miniaturized platforms.

## I. INTRODUCTION

Hydrogen ($H_2$) is widely recognized as a key energy carrier for next-generation clean energy systems, especially in fuel cell applications [1], [2]. However, its physicochemical properties, including being colorless, odorless, and highly flammable at concentrations as low as 4% in air, introduce serious safety concerns [3]. This underscores the need for robust sensing technologies capable of detecting leaks at sub-flammable levels. Fast and accurate detection is critical in confined environments such as aerospace, automotive, and industrial settings, where hydrogen accumulation can pose significant hazards [4]. Consequently, there is increasing demand for compact, sensitive, and reliable hydrogen sensors that perform well across diverse environmental conditions [5].

Recently, resonant frequency-based hydrogen sensors have gained significant attention in research [6] due to their high sensitivity, fast response time, low power requirements, and compatibility with on-chip integration. These sensors detect variations in gas concentration by monitoring shifts in the resonator's natural frequency. Palladium (Pd) is commonly used as a sensing layer due to its hydrogen absorption capability, including reversible absorption behavior as well as high selectivity over other gases [7], [8]. While alternative materials such as other noble metals (e.g., platinum), Pd alloys (e.g., Pd–Ag, Pd–Ni), metal oxides (e.g., ZnO, $SnO_2$), and conducting polymers (e.g., polyaniline) have also been explored [9], Pd remains a leading choice for applications requiring high sensitivity to hydrogen and fast response.

A variety of resonator architectures have been explored for hydrogen sensing, each leveraging frequency shifts induced by gas interactions. These include film bulk acoustic resonators (FBAR) [10], [11], surface acoustic wave (SAW) devices [3], [12]–[14], quartz crystal microbalances (QCM) [15], [16], and piezoelectric micro-diaphragm (PMD) resonators [6]. However, despite extensive research on these platforms, laterally vibrating resonators (LVRs) remain largely unexplored for hydrogen sensing applications. In this work, we present, to the best of our knowledge, the first implementation of a palladium-coated LVR for hydrogen sensing and experimentally demonstrate its high sensitivity in the low ppm range. The following sections detail the modeling, fabrication, and experimental characterization of the proposed sensor.

## II. PALLADIUM-COATED MEMS RESONATORS

### A. Working Principle

Contour-extensional mode piezoelectric MEMS resonators, also known as laterally vibrating resonators, generate high-frequency bulk acoustic waves within ultra-thin piezoelectric nanoplates. Their compact size and high quality factor make them ideal for highly sensitive and miniaturized sensing systems. A typical LVR consists of a suspended piezoelectric nanoplate (scandium-doped aluminum nitride (ScAlN) in this case), anchored at both ends and sandwiched between a bottom interdigitated electrode (IDE) and a top floating electrode. When an electrical signal is applied to the IDE, the top electrode enhances vertical field confinement, activating a high-order contour-extensional mode via the $d_{31}$ coupling coefficient of ScAlN. The resonance frequency $f_s$ is determined by the IDE pitch $W_0$, equivalent Young's modulus $E_{eq}$, and equivalent mass density $\rho_{eq}$ of the resonator stack, and is expressed as:

$$f_s = \frac{1}{W_0}\sqrt{\frac{E_{eq}}{\rho_{eq}}} \quad (1)$$

Because $f_s$ is directly determined by the pitch size, the device's operational frequency is set lithographically.

### B. Fabrication

The device shown in Fig. 1 was fabricated using a six-mask microfabrication process. Fabrication begins with patterning, sputtering, and lift-off of a 100 nm platinum bottom electrode. A 350 nm layer of 30 % Sc-doped aluminum nitride (AlN) is then deposited via reactive sputtering, followed by deposition and patterning of a 100 nm aluminum top electrode. A 250 nm AlN passivation layer is subsequently deposited by reactive sputtering. The resonant plate is then defined by ion milling through both ScAlN and AlN layers to form etch-release pits. Vias are opened by wet etching in hot phosphoric acid, and contact pads are formed by gold evaporation. Finally, a 150 nm palladium sensing layer is patterned via electron beam evaporation, and the resonator is released using isotropic $XeF_2$ etching.

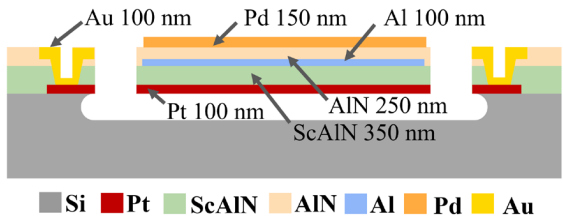

Fig. 1. Schematic cross-sectional view of the ScAlN resonator with palladium.

## C. System Modeling

When palladium is exposed to a hydrogen-containing gas mixture, the dissociated $H_2$ atoms interact with the Pd lattice, leading to volumetric expansion [17]. In LVRs, the Pd-coated stack is mechanically anchored to the substrate, preventing free expansion. This mechanical constraint induces internal compressive stress ($\sigma$) in the Pd layer, which bends the suspended resonant structure and lowers its resonant frequency. The magnitude of this frequency shift correlates directly with the hydrogen concentration. For a 128 nm-thick Pd film, compressive stress can reach up to 600 MPa within the 0–1% $H_2$ concentration range [17]. To replicate this behavior, 3D finite element simulations were carried out in COMSOL Multiphysics by sweeping the applied pre-stress in the Pd layer from 0 to 600 MPa. The induced mechanical deformation led to a downward shift in the resonant frequency, as shown in Fig. 2, confirming the LVR's sensitivity to stress changes caused by hydrogen absorption.

## III. EXPERIMENTAL RESULTS

### A. Devices Characterization

The device was characterized using a Keysight P5008A vector network analyzer (VNA) in a one-port configuration on a probe station. A Cascade ground-signal-ground (GSG) probe with a 150 µm pitch and a low-loss coaxial RF cable was used, with short-open-load (SOL) calibration ensuring accurate measurements. The extracted mechanical quality factor was $Q_m \sim 820$, and the electromechanical coupling coefficient was $k_t^2 \sim 3.18\%$. Fig. 3 shows the measured admittance around the resonance frequency $f_s = 120$ MHz, along with the fitted Modified Butterworth–Van Dyke (MBVD) model.

Following electromechanical characterization, hydrogen sensing measurements were performed. The device was wire-bonded to a PCB with an SMA connector, since the probe station could not be used under hydrogen flow. This limitation was due to space constraints and the incompatibility of the gas mixing chamber with the RF probes and manipulators used in the probe station. The device was then recharacterized with electronic calibration (ECal), ensuring measurement accuracy.

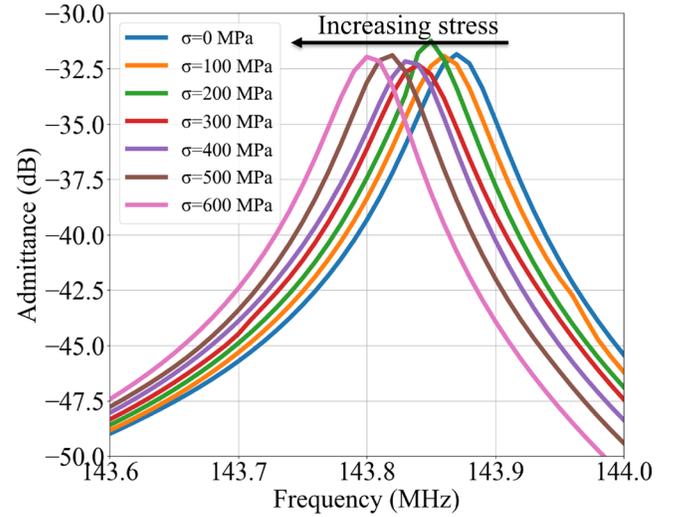

Fig. 2. Simulated resonance frequency shifts in admittance curves, zoomed around $f_s$, under varying compressive stress applied to the Pd layer.

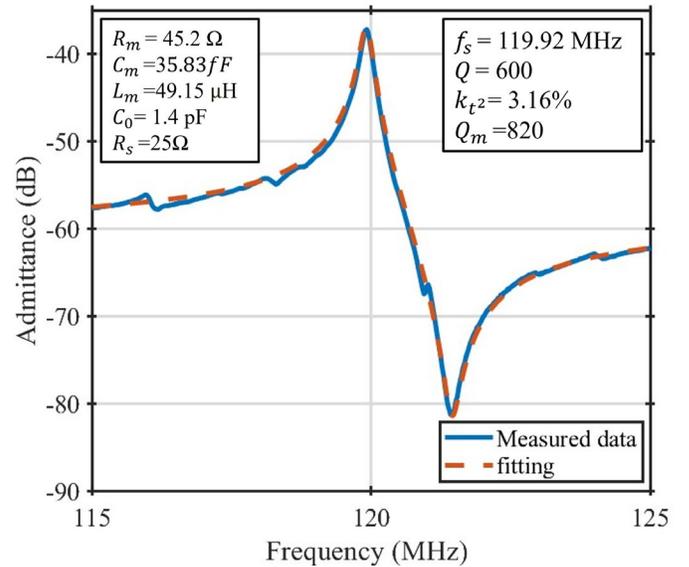

Fig. 3. Measured admittance response of the resonant sensor and the corresponding MBVD fitting.

### B. Experimental setting for Hydrogen Sensing

As shown in Fig. 4, the gas setup consists of two cylinders, one with pure nitrogen ($N_2$) and one with 1 % $H_2$ diluted in $N_2$, each controlled by mass flow controllers and merged into a custom chamber housing the PCB-mounted chip. The chamber features a gas inlet and an outlet that allows electrical connection between the PCB and the VNA. The total flow rate was fixed at 100 standard cubic centimeters per minute (sccm) to maintain stable pressure inside the chamber, yielding complete chamber venting (~0.2 L) in about 2 minutes. Two hydrogen concentration sweeps were conducted: one increasing from 0 to 10,000 ppm in 2,500 ppm steps, and the other decreasing back to 0 ppm, to assess sensor response and reversibility. Each gas concentration step was held for 20 minutes to ensure full stabilization of the $H_2$-induced stress in Pd. This duration was selected based on literature reports of 16-minute stabilization time for $\sim$128 nm Pd layer [17], with an adjustment for the 150 nm Pd layer used in this work. Fig. 5 displays the admittance curves for increasing $H_2$ concentrations, highlighting the downward resonance frequency shift induced by hydrogen absorption.

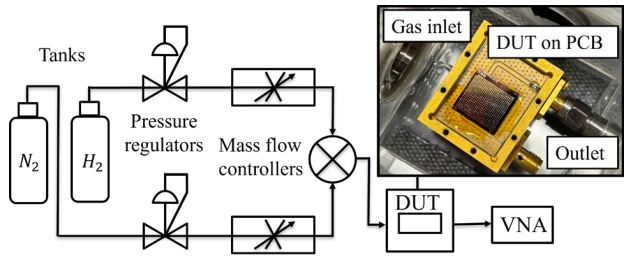

Fig. 4. Schematic of the gas test setup.

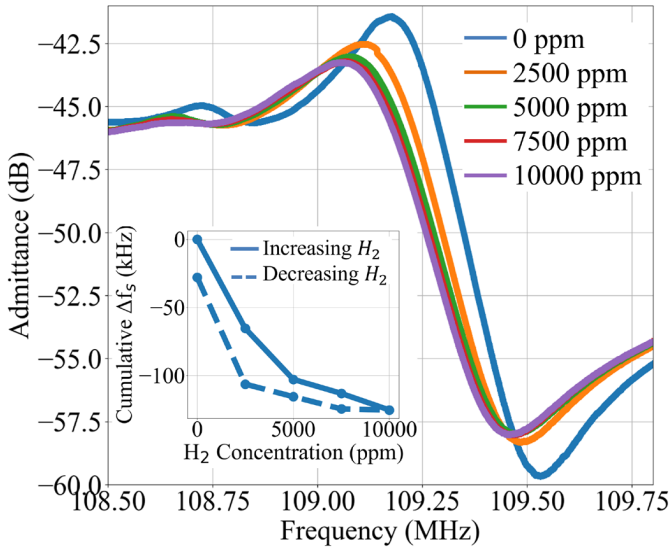

Fig. 5. Experimental admittance response of the sensor at different $H_2$ concentrations, along with the cumulative resonant frequency shifts ($\Delta f_s$) under forward and reversed sweeps.

### IV. Discussion and Future Directions

This section evaluates the performance of the Pd-coated LVR sensor. A Figure of Merit (FoM) is defined as the sensitivity normalized to the operating frequency, i.e., the resonant frequency shift per ppm of $H_2$ divided by the device's operating frequency (expressed in Hz/MHz/ppm). Table I summarizes the FoM, quality factor ($Q$), lateral dimensions, and lithographic frequency definition. Our device outperforms comparable MHz-range technologies, offering up to 50$\times$ higher sensitivity, the highest $Q$, the smallest lateral footprint, and multi-frequency definition enabled by a single lithographic step. Moreover, our device exhibits a FoM comparable to higher-frequency FBARs [10], [11], while providing lithographic definition, higher $Q$, and a fivefold reduction in size. The PMD device [6] achieves the highest FoM, but has a lower $Q$ and is six times larger.

Further measurements are needed to establish the reversibility and linearity of the sensor. As shown in Fig. 5, increasing $H_2$ concentration in the Pd layer causes a downward shift in the resonant frequency, with the largest drop (66 kHz) occurring between 0 and 2,500 ppm, corresponding to a sensitivity of about 26 Hz/ppm. However, at higher concentrations, the response becomes nonlinear and gradually saturates. This saturation may result from residual stress and limited mechanical compliance in the multilayer stack, which restricts further expansion of the Pd layer. From the reversed $H_2$ sweep, we observed that the sensor response exhibited partial reversibility, with the resonance frequency not entirely returning to its initial value. This suggests incomplete $H_2$ desorption from the Pd layer, leaving residual stress that keeps the resonance slightly downshifted. Mitigation strategies under investigation include geometry optimization and stack refinement, i.e., layer thinning and AlN passivation removal, and integrating a heater pad accessible via DC probes, to facilitate hydrogen desorption from the Pd layer through localized heating [18], [19].

### V. Conclusion

This work demonstrates the potential of MEMS laterally vibrating resonators with palladium coatings for miniaturized hydrogen sensing, achieving detection up to 10,000 ppm and a peak sensitivity of 26 Hz/ppm in the low ppm range. Further improvements in sensitivity, response time, and quality factor are expected through optimization of the device geometry. Nonlinear saturation at higher concentrations and limited reversibility are under investigation and may be addressed through improved mechanical compliance through stress optimization and integration of heating pads to facilitate hydrogen desorption.

| FoM $\frac{Hz}{MHz \cdot ppm}$ | $Q$ | Freq. (MHz) | Lat. Dim. ($\mu$m) | Litho. Def. | Device Type |
|---|---|---|---|---|---|
| 0.01 | LOW | 102 | 5000 | YES | SAW [3] |
| 0.04 | LOW | 129 | 6000 | YES | SAW [13] |
| 0.004 | LOW | 70 | 3200 | YES | SAW [12] |
| 0.01 | LOW | 160 | 3200 | YES | SAW [14] |
| 0.2 | HIGH | 2230 | 500 | NO | FBAR [10] |
| 0.2 | HIGH | 2390 | 500 | NO | FBAR [11] |
| 0.003 | MEDIUM | 165 | 1800 | NO | QCM [15] |
| 0.00003 | MEDIUM | 9 | 5000 | NO | QCM [16] |
| 12 | LOW | 0.15 | 600 | YES | PMD [6] |
| 0.24 | HIGHER | 109 | 100 | YES | This work* |

TABLE I
Comparison of Resonant Hydrogen Sensors.

### VI. Acknowledgments

The authors thank the staff at the George J. Kostas Nanoscale Technology and Manufacturing Research Center at Northeastern University for their support. This work was funded by the ARPA-E OPEN 2025 program.
Gaia Giubilei and Farah Ben Ayed contributed equally to this work.